\documentclass[final]{raa06}           
\usepackage{graphicx,times}             
\usepackage{natbib}
\usepackage{amssymb,amsmath}
\bibpunct{(}{)}{;}{a}{}{,}
\usepackage[colorlinks=true, citecolor=blue]{hyperref}%

\usepackage{graphicx,kantlipsum,setspace}
\usepackage{caption}
\usepackage{graphics,epsf}
\usepackage{amsmath}                
\usepackage{amsfonts}               
\usepackage{amssymb}                
\usepackage{epsfig}                 
\usepackage{appendix}
\usepackage{graphicx}
\usepackage{float}
\usepackage{color}
\usepackage{multirow}
\usepackage{colortbl}
\usepackage[para,online,flushleft]{threeparttable}
\usepackage{xcolor}

\hypersetup{citecolor=blue, 
            linkcolor=red, 
            menucolor=blue, 
            urlcolor=blue}  

\newcommand{\cm}{{~\rm cm}}

\newcommand{\km}{{~\rm km}}
\newcommand{\s}{{~\rm s}}

\newcommand{\erg}{{~\rm erg}}
\newcommand{\yr}{{~\rm yr}}

\begin{document}
\titlerunning{A DD-MED scenario to explain SN Ia 2020aeuh}
   \title{A double-degenerate scenario with a merger to explosion delay time to explain type Ia supernova SN 2020aeuh 
}

   \volnopage{Vol.0 (20xx) No.0, 000--000}      
   \setcounter{page}{1}          


   \author{Noam Soker
    }

   \institute{Department of Physics, Technion - Israel Institute of Technology, Haifa, 3200003, Israel;   {\it    soker@physics.technion.ac.il} \\ 
   \vs\no
   {\small Noam Soker: orcid: {0000-0003-0375-8987}}\\
\vs\no
   {\small Received~~20xx month day; accepted~~20xx~~month day}}

\abstract{
I suggest the double-degenerate (DD) scenario with a merger-to-explosion delay (MED) time (the DD-MED scenario) of about 1-2 years to explain the rare properties of the recently analyzed type Ia supernova (SN Ia) SN 2020aeuh. The rare properties are the SN Ia ejecta interacting with a carbon-oxygen (CO)-rich circumstellar material (CSM) at approximately 50 days post-explosion. In this DD-MED scenario, two massive CO white dwarfs (WDs), masses of $M_1 \simeq 1.1 M_\odot$ and $M_2 \simeq 1 M_\odot$, merge to leave a rapidly rotating lonely WD of about the Chandrasekhar mass. The merger process ejects $M_{\rm CSM} \simeq 0.7 M_\odot$ to form a nonspherical CO-rich CSM.  At the explosion, there is a lonely WD and a detached hydrogen- and helium-deficient CSM. Studies proposed the other lonely WD scenario, the core-degenerate (CD) scenario, to explain several specific SNe Ia and SN Ia remnants. SN 2020aeuh is the first particular SN Ia that is attributed to the DD-MED scenario. Besides being slightly brighter than typical SNe Ia and the CSM interaction, SN 2020aeuh is a normal SN Ia. Therefore, this study strengthens the claim of earlier studies, which are based on other arguments, like the properties of SN Ia remnants,  that the lonely WD scenarios, i.e., the DD-MED and CD scenarios, might account for most, if not all, normal SNe Ia. These earlier studies also argue that  all SN Ia scenarios, whether lonely WD or not, might contribute to peculiar SNe Ia. 
 }

\maketitle

\section{INTRODUCTION}
\label{sec:intro}

The old classification of type Ia supernovae (SNe Ia) scenarios, into the double-degenerate (DD) and single-degenerate (SD) scenarios, is outdated (despite many papers still mentioning this classification as if it is applicable). There are more scenarios that any modeling of an SN Ia must consider. However, there is no consensus on the classification of these scenarios as reviews show (\citealt{Maozetal2014, MaedaTerada2016, Hoeflich2017, LivioMazzali2018, Soker2018Rev, Soker2019Rev, Soker2024Rev, Wang2018,  Jhaetal2019NatAs, RuizLapuente2019, Ruiter2020, Aleoetal2023, Liuetal2023Rev, Vinkoetal2023, RuiterSeitenzahl2025}). No scenario is entirely free of difficulties in accounting for observations, with some encountering a few challenges, while others face many challenges. Some observations challenge most (e.g., \citealt{Pearsonetal2024}) or all (e.g., \citealt{Wangetal2024}) scenarios. A challenging observation, for example,
is the relation of the $\gamma$-ray escape time to the $^{56}$Ni mass (e.g., \citealt{SchinasiLembergKushnir2025, sharonKushnirWygoda2025, Sharonetal2025}). 
Studies have added more sub-scenarios or branches of the main scenarios, including the HeCO hybrid channel of the DD scenario (e.g., \citealt{Peretsetal2019, Zenatietal2019, Zenatietal2023} ) common-envelope wind model \citep{MengPodsiadlowski2017, CuiMeng2022, WangMeng2025}, the carbon–oxygen–neon white dwarf (WD) with a main sequence companion channel of the SD scenario (e.g., \citealt{GuoMeng2025RAA}), the core-merger detonation model \citep{Ablimit2021}, and new mass transfer prescriptions (e.g., \citealt{Lietal2023RAA}). Studies have examined most of these scenarios in recent years, with no emerging leading scenario(s) in the consensus (e.g., \citealt{Boraetal2024, Bossetal2024, Bregmanetal2024, CasabonaFisher2024, DerKacyetal2024, Joshietal2024, Koetal2024, Kobashietal2024, Limetal2024, Mehtaetal2024, Palicioetal2024, Phillipsetal2024, Soker2024RAAPN, Uchidaetal2024, Burmesteretal2025, Courtetal2025, Gabaetal2025NewA, Glanzetal2025, Griffithetal2025, Hoogendametal2025a, Hoogendametal2025b, Itoetal2025, IwataMaeda2025, Kumaretal2025, Mageeetal2025, MichaelisPerets2025, OHoraetal2025, Panetal2025, Pollinetal2025, Simotasetal2025, WangChenPan2025}, for a limited list of papers since 2024). 

In some cases, papers debate the best scenario for a specific SN Ia. A recent example is the debate over the progenitor of supernova remnant (SNR) 0509-67.5, where  \cite{Dasetal2025NatAs} argue for the double detonation (DDet) scenario. In contrast, I, \citep{Soker2025SNR0509}, argued for the core-degenerate (CD) scenario. In general, despite its popularity among many researchers (e.g., \citealt{Callanetal2024, MoranFraileetal2024, PadillaGonzalezetal2024, Polinetal2024, Shenetal2024, Zingaleetal2024, Rajaveletal2025, Wuetal2025}), the DDet scenario encounters challenges in explaining normal SNe Ia (e.g., \citealt{BraudoSoker2024, BraudoSoker2025RAA, Soker2024Comment, Soker2025SNR0509}) but might be more relevant to peculiar SNe Ia. As well, some runaway WDs might result from type Iax SNe (e.g., \citealt{Igoshevetal2023}), a group of peculiar SNe Ia, rather than the DDet scenario for normal SNe Ia with a degenerate companion (DDet-DD scenario).   

In addition to the classification of scenarios, there is the classification by the ignition process, like an outer helium detonation in the DDet scenario and its channels, or the deflagration to detonation transition in Chandrasekhar mass WDs, and the classification by the WD mass, sub-Chandrasekhar or near Chandrasekhar ($M_{\rm Ch}$); see the above listed reviews. The classification most relevant to this study is the one that distinguishes between lonely WD scenarios and non-lonely ones. In this classification, normal SNe Ia are descendants of lonely WD scenarios, while peculiar SNe Ia might result from all scenarios \citep{Soker2024Rev}. 

The lonely WD scenarios involve pre-explosion binary interaction, where at the time of explosion, there is only one WD, with little or no leftover from the binary interaction bound to the lonely WD; there might be a circumstellar material (CSM), and even a massive CSM, but it is expanding and not bound. The WD does not survive the explosion in normal SNe Ia, and might survive in some peculiar SNe Ia. The WD is a merger product of a core of an asymptotic giant branch star with an older WD in the CD scenario, or the merger of two WDs in the DD scenario with a merger-to-explosion delay (MED) time, the DD-MED scenario. The MED time must be much longer than the dynamical time of the merger to allow the WD remnant of the merger to relax and lead to a more or less spherical explosion. The notion of MED time in SN Ia scenarios is several years old (e.g., \citealt{Soker2018Rev, Soker2022RAA, Neopaneetal2022}), while the idea of the group of lonely WD scenarios is two years old \citep{Soker2024Rev}. The CD scenario has a MED time built in; studies have applied it to several SN Ia remnants (e.g., SNR G1.9+0.3, \citealt{Soker2024RAAPN}; SNR 0509-67.5, \citealt{Soker2025SNR0509}) and to the SN Ia PTF 11kx \citep{Sokeretal2013}, which interacted with CSM. In this study, I argue that the DD-MED scenario is the most suitable one to explain SN 2020aeuh, which \cite{Tsalapatasetal2025} recently analyzed and found to be enigmatic. To the best of my knowledge, this is the first specific SN Ia to which the DD-MED scenario in the frame of a lonely WD explosion is applied.

\section{The DD-MED scenario for SN 2020aeuh}
\label{sec:DDMED}

SN 2020aeuh \citep{Tsalapatasetal2025} is a bright SN Ia with a delayed interaction with a carbon-oxygen (hydrogen and helium deficient) CSM. \cite{Tsalapatasetal2025} study its properties in detail and discuss in length the problems with many existing SN Ia scenarios. However, they do not discuss the DD-MED scenario at all. In the DD-MED scenario, two CO WDs merge, and the explosion occurs after the lonely WD remnant of the merger has time to relax dynamically; this time is the MED time. I argue that the DD-MED scenario provides the best explanation for SN 2020aeuh.  

In Table \ref{Tab:Table1}, I summarize the main properties of SN 2020aeuh (first column), the way the DD-MED scenario accounts for each property (second column; for reviews on the DD-MED scenario see  \citealt{Soker2019Rev} and \citealt{Soker2024Rev}), and the problems of other scenarios (third column). 
\begin{table*}
\begin{center}
  \caption{Properties of SN 2020aeuh and the DD-MED scenario explanation}
    \begin{tabular}{| p{3.6cm} | p{7.0cm}| p{5.0cm}| }
\hline  
\textbf{Property} & \textbf{DD-MED scenario} & \textbf{Challenges for other scenarios}  \\
\hline  
Hydrogen deficient CSM& The merger of two CO WDs with mass ejection  & Contradicts the SD scenario  \\ 
\hline  
Helium deficient CSM & The merger of two CO WDs with mass ejection  & Contradicts the DDet scenario  \\ 
\hline  
$M_{\rm CSM} \simeq 1 M_\odot$& Merger of $M_1\simeq 1.1$ and $M_2 \simeq 1M_\odot$ leaves a lonely WD remnant of $M_{\rm lonely} \simeq 1.4 M_\odot$ &  Contradicts the WWC scenario \\ 
\hline  
Ejecta-CSM interaction $\simeq 50~$days after explosion & The MED time was $t_{\rm MED} \simeq 1-2\yr$  & Contradicts the DD scenario where the explosion has no MED; challenges the DDet scenario  \\ 
\hline  
Abrupt rise and then longer drop in the mass loss of the CSM$^\ast$ & Mass transfer with mass loss started in a short time at the beginning of binary interaction, declined over a longer period until full merger  &   \\ 
\hline  
$v_{\rm CSM} \lesssim 1500 \km \s^{-1} $ &  Main mass loss from the second Lagrange point  &      \\
\hline  
No very late radio emission & There was no mass loss event before the binary merger process  & A challenge to the CD scenario, unless new ejecta-CSM interaction occurs in $\approx 1000 \yr$ \\
\hline  
The host galaxy shows no star formation & The WD binary is old and brought to the merger by gravitational waves & A challenge to the CD scenario, unless some star formation is found to have occurred in the last $\simeq 10^8 \yr$     \\
\hline  
     \end{tabular}
  \label{Tab:Table1}\\
\end{center}
\begin{flushleft}
\small 
Notes: First column: the main properties of SN 2020aeuh as I compiled from \cite{Tsalapatasetal2025}. Second column: The explanation according to the DD-MED scenario. Third column: The challenges of other SNe Ia scenarios. 
\newline
$^\ast$ A property that \cite{Tsalapatasetal2025} deduce from their interaction model and not directly from observables.
\newline
Abbreviations: CD: core degenerate;  DD: double degenerate; DDet: double detonation; SD: single degenerate; WWC: WD-WD collision, a scenario where two unbound WDs collide; 
$M_{\rm CSM}$: The CSM mass; $M_{\rm lonely}$: the mass of the lonely WD remnant of the merger; $v_{\rm CSM}$: the expansion velocity of the CSM before interacting with the SN ejecta; MED: merger to explosion delay. 
\end{flushleft}
\end{table*}

In the DD-MED scenario that I propose for SN 2020aeuh, two massive CO WDs, of masses $M_1 \simeq 1.1 M_\odot$ and $M_2\simeq 0.95-1.1 M_\odot$, merge to form a rapidly rotating lonely WD at about the Chandrasekhar mass, $M_{\rm lonely} \simeq 1.4 M_\odot$; at explosion, there is only one WD, the lonely WD remnant. \cite{IlkovSoker2012} and \cite{Neopaneetal2022} discuss the justifications and arguments for the MED time of years and more.  

The ejected mass is then $M_{\rm CSM} \simeq 0.65 - 0.8 M_\odot$. \cite{Tsalapatasetal2025} estimates the CSM mass in a spherical shell to be $\simeq 1-2 M_\odot$ and comment that a non-spherical distribution might allow a lower CSM mass. If the CSM covers a fraction of $\beta \simeq 0.5$ of the sphere, then a fraction $\beta$ of the ejecta interacts with the CSM; the interacting ejecta and the CSM have a similar mass for these parameters. As the ejecta is much faster than the CSM, about an order of magnitude faster, $v_{\rm ej} \simeq 1.2 \times 10^4 \km \s^{-1}$ and $v_{\rm CSM} \simeq 1200 \km \s^{-1}$, momentum conservation implies that the two media expand at about half the ejecta velocity. The interaction converts about half of the kinetic energy of the interacting ejecta to heat. This amounts to $\simeq 0.25$ of the total ejecta kinetic energy. If a fraction $\xi_{\rm rad} \simeq 0.5$ of this thermal energy is radiated, I find $E_{\rm rad} \simeq 0.12 E_{\rm kin}$, where $E_{\rm kin}$ is the kinetic energy of the entire SN ejecta. I could also take a conversion factor of $\epsilon \simeq 0.3$ of the kinetic energy of the interacting ejecta to radiation, as \cite{Tsalapatasetal2025} do; this will give $E_{\rm rad} \simeq 0.15 E_{\rm kin}$. The kinetic energy of this bright SN Ia might be somewhat larger than $E_{\rm kin} =10^{51} \erg$. Overall, the CSM mass of $M_{\rm CSM} \simeq 0.7 M_\odot$ can account for the radiated energy of $E_{\rm rad} \simeq 1.1 \times 10^{50} \erg$, that \cite{Tsalapatasetal2025} calculate for SN 2020aeuh. More accurate treatment is needed for the proposed non-spherical CSM distribution.  

\cite{Tsalapatasetal2025} find the full width at half-maximum intensity of two narrow (relative to the ejecta) spectral lines to be $\simeq 750 \km \s^{-1}$ and $\simeq 1250 \km \s^{-1}$, and a third asymmetrical line shows a velocity of $\simeq 1350 \km \s^{-1}$. I, therefore, scale the CSM velocity with $v_{\rm CSM} \simeq 1200 \km \s^{-1}$. \cite{Tsalapatasetal2025} take the ejecta velocity to be $v_{\rm ej} \simeq 1.2 \times 10^4 \km \s^{-1}$, and from that estimate the CSM shell to be from $R_{\rm CSM, inner}\simeq 3 \times 10^{15} \cm$ to $R_{\rm CSM, outer}\simeq 9 \times 10^{15} \cm$.  The CSM velocity is about an order of magnitude slower. From the explosion until the ejecta catches up with the CSM, the radius of the CSM increases by a factor of $\simeq 1.1$, i.e., by $\simeq 10\%$. Therefore, for the accuracy of this calculation, the exact expansion of the CSM from explosion to the time of ejecta-CSM interaction is of little significance.  For a median CSM velocity of $v_{\rm CSM} \simeq 1200 \km \s^{-1}$, the time from CSM ejection to explosion, namely, the MED time, is $t_{\rm MED} \simeq 0.5 (R_{\rm CSM, inner} + R_{\rm CSM, outer})/v_{\rm CSM}$.  The MED time from the merger event to the explosion is, therefore,  $t_{\rm MED} \simeq 6\times 10^{15} \cm /1,200 \km \s^{-1} = 5 \times 10^7 \s$, and so I estimate $t_{\rm MED} \simeq 1-2 \yr$. The value of $v_{\rm CSM}$ is the average velocity of the CSM. The inner parts of the CSM are slower. For a delay time of $t_{\rm MED} \simeq1.5 \yr$, an inner CSM velocity of $v_{\rm CSM,inner} \simeq 650 \km \s^{-1}$ gives the observed inner CSM radius of $R_{\rm CSM, inner,D}  \simeq v_{\rm CSM,inner} t_{\rm MED} \simeq 3.1 \times 10^{15} \cm$. This is the DD-MED explanation for the inner CSM radius.    

During the merger event, the binary system ejects the CSM. The escape velocity from the merger system, $\approx 5000 \km \s^{-1}$, as \cite{Tsalapatasetal2025} note, is much larger than the CSM expansion velocity $v_{\rm CSM} \simeq 1200 \km \s^{-1}$. I suggest that most of the CSM mass was lost in and near the equatorial plane through the second Lagrange point, i.e., on the side of the less massive WD. Mass lost from the second Lagrange point in strong binary interaction might be at a terminal velocity much slower than the escape velocity from the system, or even stays bound (e.g, \citealt{HubovaPejcha2019} and references therein).  
\cite{RaskinKasen2013} simulated the merger of two WDs in the DD scenario, and found a tidal tail expanding at $\approx 2000 \km \s^{-1}$. They found the ejected mass to be $\approx 0.001-0.005 M_\odot$. The lonely WD scenario lets the merger product relax and eject more mass.   

Other properties of SN 2020aeuh and how the CD-MED scenario accounts for them are evident from Table \ref{Tab:Table1}, and I further discuss some of these properties in Section \ref{sec:Problems}.  

\section{The problems with some other scenarios}
\label{sec:Problems}

\cite{Tsalapatasetal2025} discussed in length some problems with some SN Ia scenarios, and I will not repeat their discussions. Here, I focus on the problems with other scenarios in comparison with the success of the DD-MED scenario in explaining SN 2020aeuh, because  \cite{Tsalapatasetal2025} did not discuss the DD-MED scenario. I follow the third column of Table \ref{Tab:Table1} and refer only to the properties of SN 2020aeuh; in Section \ref{sec:Summary}, I will discuss some SN Ia scenarios concerning the general population of normal and peculiar SNe Ia. 

The hydrogen-deficient CSM rules out the SD scenario for SN 2020aeuh, as the SD scenario can only explain hydrogen-rich CSM. 

The DDet scenario, where one star transfers helium-rich gas to a CO WD, predicts helium-rich CSM, if at all. Additionally, it is unclear whether the DDet scenario can account for the delay between the mass loss episode that formed the CSM and the explosion, as the DDet generally has no MED time (MED time also refers to time from a mass transfer episode to the explosion).  

The WD-WD collision scenario, where two unbound WDs collide and ignite on a dynamical timescale, predicts no close CSM, independent of the composition. It cannot explain SN 2020aeuh. 

The CD scenario, which incorporates a MED as an essential ingredient, faces two challenges in explaining the properties of SN 2020aeuh; however, these properties do not completely rule it out for SN 2020aeuh. The first is that the common envelope evolution that brings the old WD to merge with the core of an asymptotic giant branch star ejects a hydrogen-rich envelope. The WD-core merger might occur in rare cases later, if the WD survives and the helium envelope expands much later to engulf the WD, causing it to merge with the core.  \cite{SokerBear2023} suggested this channel of the CD scenario to explain the compact helium-rich CSM of SN 2020eyj \citep{Kooletal2023}. However, it is unclear if a common envelope evolution can take place without helium presence in the ejected CSM. According to the CD scenario, the ejecta will eventually collide, within approximately a thousand years, with a hydrogen-rich nebula that was once a planetary nebula. According to the CD scenario, the asymptotic giant branch that engulfs the WD should have a zero-age main-sequence mass of $\gtrsim 4 M_\odot$ to force the WD to merge with the core. Thus, some relatively recent star formation must have taken place in the host galaxy of SN 2020aeuh (in the last $\simeq 10^8 \yr$). These two challenges of the CD scenario make the DD-MED scenario the most suitable for explaining SN 2020aeuh.

\section{Summary}
\label{sec:Summary}

I argued (Section \ref{sec:DDMED}) that the DD-MED scenario best explains SN 2020aeuh that \cite{Tsalapatasetal2025} recently analyzed and found to be enigmatic. The merger of two massive CO WDs that form a lonely WD remnant that explodes with a MED time of $t_{\rm MED} \simeq 1-2 \yr$ can account for all the properties of SN 2020aeuh (Table \ref{Tab:Table1}). All other scenarios encounter difficulties, as noted by \cite{Tsalapatasetal2025}, and I list them in the third column of Table \ref{Tab:Table1}. 

The DD-MED and CD scenarios form the group of lonely WD SN Ia scenarios \citep{Soker2024Rev}, where a massive lonely WD that is a remnant of a merger event explodes as a SN Ia.   
The CD scenario has a MED time built in; studies have applied it to several SN Ia remnants (e.g., SNR G1.9+0.3, \citealt{Soker2024RAAPN}; SNR 0509-67.5, \citealt{Soker2025SNR0509}) and to the SN Ia PTF 11kx \citep{Sokeretal2013}, which interacted with a CSM. \cite{Tsalapatasetal2025} found SN 2020aeuh to have some similarities with two other SN Ia-CSM, PTF 11kx and SN 2020eyj, e.g., in the light curve until about 25 days after peak brightness. Studies proposed the CD scenario for these two SNe Ia (\citealt {Sokeretal2013} for PTF 11kx and \citealt{SokerBear2023} for SN 2020eyj), where each of these two SNe Ia had a lonely WD exploding inside an old planetary nebula. This study is the first to account for a specific SN Ia within the CD-MED scenario, the other lonely WD scenario. 

SN 2020aeuh belongs to the group of somewhat overluminous SNe Ia, specifically 1991T-like SNe Ia, that some take to be super-Chandrasekhar (e.g., \citealt{Fisheretal1999}), namely, result from a WD with a mass $>M_{\rm Ch}$. Often SNe Ia-CSM are overluminous (1991T-like SNe Ia; e.g., \citealt{Phillipsetal2024}), but otherwise are normal SNe Ia. In the lonely WD scenarios, what separates normal SNe Ia that interact with a close CSM from other normal SNe Ia is a relatively short MED time. I raise the possibility that the reason for the short MED time of SNe Ia-CSM and their overluminous brightness is the mass of the exploding lonely WD being on the upper end of the mass distribution of exploding lonely WDs. Namely, they are indeed somewhat super-Chandrasekhar.     

I crudely estimate the expected number of SNe Ia that interact with a CO-rich CSM as follows. The MED maximum time can be as long as $t_{\rm MED,M} \approx 3 \times 10^6 \yr$ with possibly constant rate (\citealt{Soker2019b}). The fraction of the DD-MED scenario is less than half, as I expect most SNe Ia to come from the CD scenario (\citealt{Soker2022RAA}); I will scale with $\alpha_{\rm DM} \approx 0.5$. Most of the DD-MED merger will result in a Chandrasekhar mass white dwarf without much leftover, namely, $ M_1+M_2 \simeq 1.4 M_\odot$. Only a small fraction of more massive WDs eject a massive CSM. I scale with $\alpha_{\rm Ch} \approx 0.2$.  For a SN Ia ejecta ten times as fast as the merger ejecta, the expected fraction of SNe Ia with ejecta interaction with a CO-rich CSM within a post-explosion time of $\Delta t_{\rm pe}$ is 
\begin{equation}
f \approx 10^{-5}  
\left( \frac{\alpha_{\rm DM}}{0.5}  \right)
\left( \frac{\alpha_{\rm Ch}}{0.2}  \right)
\left( \frac{t_{\rm MED,M}}{10^6 \yr} \right)^{-1}
\left( \frac{\Delta t_{\rm pe}}{10 \yr} \right). 
\label{eq:Fraction}
\end{equation}
I scale with $\Delta t_{\rm pe} = 10 \yr$, which is the typical time scale that observations are searching for late ejecta-CSM interaction after the SN Ia explosions  
The uncertainties are very large, more than an order of magnitude, and the only certain point is that SNe Ia-CSM-CO are very rare; all types of SNe Ia-CSM are rare (e.g., \citealt{Terweletal2025a, Terweletal2025b}).  Better estimates of the fraction of such events are for the future.

SN 2020aeuh supports the suggestion that the outcomes of WD mergers are not only SNe Ia or massive WDs that live forever (e.g., \citealt{Yangetal2022}), but also massive lonely WDs that eventually explode as SNe Ia, i.e., the DD-MED scenario. With that, SN 2020aeuh strengthens the suggestion that lonely WDs are responsible for most (or all) normal SNe Ia. All other scenarios might contribute to peculiar SNe Ia (e.g., \citealt{Soker2024Rev}).

\section*{Acknowledgments}
I thank an anonymous referee for useful comments. I thank the Charles Wolfson Academic Chair at the Technion for the support.






\begin{thebibliography}{}

\bibitem[\protect\citeauthoryear{Ablimit}{2021}]{Ablimit2021} Ablimit I., 2021, PASP, 133, 074201. 

\bibitem[\protect\citeauthoryear{Aleo et al.}{2023}]{Aleoetal2023} Aleo P.~D., Malanchev K., Sharief S., Jones D.~O., Narayan G., Foley R.~J., Villar V.~A., et al., 2023, ApJS, 266, 9. 

\bibitem[\protect\citeauthoryear{Boos, Townsley, \& Shen}{2024}]{Bossetal2024} Boos S.~J., Townsley D.~M., Shen K.~J., 2024, ApJ, 972, 200. 

\bibitem[\protect\citeauthoryear{Bora et al.}{2024}]{Boraetal2024} Bora Z., K{\"o}nyves-T{\'o}th R., Vink{\'o} J., B{\'a}nhidi D., B{\'\i}r{\'o} I.~B., Bostroem K.~A., B{\'o}di A., et al., 2024, PASP, 136, 094201. 

\bibitem[\protect\citeauthoryear{Braudo \& Soker}{2024}]{BraudoSoker2024} Braudo J., Soker N., 2024, OJAp, 7, 7. 

\bibitem[\protect\citeauthoryear{Braudo \& Soker}{2025}]{BraudoSoker2025RAA} Braudo J., Soker N., 2025, RAA, 25, 065012. 

\bibitem[\protect\citeauthoryear{Bregman et al.}{2024}]{Bregmanetal2024} Bregman J.~N., Gnedin O.~Y., Seitzer P.~O., Qu Z., 2024, ApJL, 968, L6. 


\bibitem[\protect\citeauthoryear{Burmester et al.}{2025}]{Burmesteretal2025} Burmester U.~P., Ferrario L., Pakmor R., Seitenzahl I.~R., 2025, MNRAS.tmp. 


\bibitem[\protect\citeauthoryear{Callan et al.}{2025}]{Callanetal2024} Callan F.~P., Collins C.~E., Sim S.~A., Shingles L.~J., Pakmor R., Srivastav S., Pollin J.~M., et al., 2025, MNRAS, 539, 1404. 

\bibitem[\protect\citeauthoryear{Casabona \& Fisher}{2024}]{CasabonaFisher2024} Casabona G.~O., Fisher R.~T., 2024, ApJL, 962, L31. 

\bibitem[\protect\citeauthoryear{Court et al.}{2025}]{Courtetal2025} Court T., Badenes C., Lee S.-H., Patnaude D., Bravo E., 2025, arXiv:2507.20828

\bibitem[\protect\citeauthoryear{Cui et al.}{2022}]{CuiMeng2022} Cui Y., Meng X., Podsiadlowski P., Song R., 2022, A\&A, 667, A154. 

\bibitem[\protect\citeauthoryear{Das et al.}{2025}]{Dasetal2025NatAs} Das P., Seitenzahl I.~R., Ruiter A.~J., R{\"o}pke F.~K., Pakmor R., Vogt F.~P.~A., Collins C.~E., et al., 2025, NatAs.tmp. 

\bibitem[\protect\citeauthoryear{DerKacy et al.}{2024}]{DerKacyetal2024} DerKacy J.~M., Ashall C., Hoeflich P., Baron E., Shahbandeh M., Shappee B.~J., Andrews J., et al., 2024, ApJ, 961, 187. 


\bibitem[\protect\citeauthoryear{Fisher et al.}{1999}]{Fisheretal1999} Fisher A., Branch D., Hatano K., Baron E., 1999, MNRAS, 304, 67. 

\bibitem[\protect\citeauthoryear{Gaba et al.}{2025}]{Gabaetal2025NewA} Gaba J., Thakur R.~K., Sharma N., Verma D., Gupta S., 2025, NewA, 120, 102411. 

\bibitem[\protect\citeauthoryear{Glanz et al.}{2025}]{Glanzetal2025} Glanz H., Perets H.~B., Bhat A., Pakmor R., 2025, arXiv:2410.17306. 


\bibitem[\protect\citeauthoryear{Griffith et al.}{2025}]{Griffithetal2025} Griffith O., Showerman G., Sarbadhicary S.~K., Harris C.~E., Chomiuk L., Sollerman J., Lundqvist P., et al., 2025, arXiv:2506.19071. 


\bibitem[\protect\citeauthoryear{Guo et al.}{2025}]{GuoMeng2025RAA} Guo B., Meng X., Tian Z., Luo J., Liu Z., 2025, RAA, 25, 015018. 

\bibitem[\protect\citeauthoryear{Hoeflich}{2017}]{Hoeflich2017} Hoeflich P., 2017, in Handbook of Supernovae, Springer International Publishing AG, 2017, p. 1151 

\bibitem[\protect\citeauthoryear{Hoogendam et al.}{2025a}]{Hoogendametal2025a} Hoogendam W.~B., Ashall C., Jones D.~O., Shappee B.~J., Tucker M.~A., Huber M.~E., Auchettl K., et al., 2025, ApJ, 988, 209. 

\bibitem[\protect\citeauthoryear{Hoogendam et al.}{2025b}]{Hoogendametal2025b} Hoogendam W.~B., Jones D.~O., Ashall C., Shappee B.~J., Foley R.~J., Tucker M.~A., Huber M.~E., et al., 2025b, arXiv, arXiv:2502.17556. 

\bibitem[\protect\citeauthoryear{Hubov{\'a} \& Pejcha}{2019}]{HubovaPejcha2019} Hubov{\'a} D., Pejcha O., 2019, MNRAS, 489, 891. 

\bibitem[\protect\citeauthoryear{Igoshev, Perets, \& Hallakoun}{2023}]{Igoshevetal2023} Igoshev A.~P., Perets H., Hallakoun N., 2023, MNRAS, 518, 6223. 

\bibitem[\protect\citeauthoryear{Ilkov \& Soker}{2012}]{IlkovSoker2012} Ilkov M., Soker N., 2012, MNRAS, 419, 1695. 

\bibitem[\protect\citeauthoryear{Ito et al.}{2025}]{Itoetal2025} Ito D., Sano H., Nakazawa K., Mitsuishi I., Fukui Y., Sudou H., Takaba H., 2025, ApJ, 978, 123. 

\bibitem[\protect\citeauthoryear{Iwata \& Maeda}{2025}]{IwataMaeda2025} Iwata K., Maeda K., 2025, ApJ, 987, 21. 

\bibitem[Jha et al.(2019)]{Jhaetal2019NatAs} Jha, S.~W., Maguire, K., \& Sullivan, M.\ 2019, Nature Astronomy, 3, 706

\bibitem[\protect\citeauthoryear{Joshi, Strolger, \& Zenati}{2024}]{Joshietal2024} Joshi B.~A., Strolger L.-G., Zenati Y., 2024, ApJ, 974, 15. 

\bibitem[\protect\citeauthoryear{Ko et al.}{2024}]{Koetal2024} Ko T., Suzuki H., Kashiyama K., Uchida H., Tanaka T., Tsuna D., Fujisawa K., et al., 2024, ApJ, 969, 116. 

\bibitem[\protect\citeauthoryear{Kobashi et al.}{2024}]{Kobashietal2024} Kobashi R., Lee S.-H., Tanaka T., Maeda K., 2024, ApJ, 961, 32. 

\bibitem[\protect\citeauthoryear{Kool et al.}{2023}]{Kooletal2023} Kool E.~C., Johansson J., Sollerman J., Mold{\'o}n J., Moriya T.~J., Mattila S., Schulze S., et al., 2023, Natur, 617, 477. 

\bibitem[\protect\citeauthoryear{Kumar, Prust, \& Bildsten}{2025}]{Kumaretal2025} Kumar G., Prust L.~J., Bildsten L., 2025, arXiv:2507.19722

\bibitem[\protect\citeauthoryear{Li, Liu, \& Wang}{2023}]{Lietal2023RAA} Li L.-H., Liu D.-D., Wang B., 2023, RAA, 23, 075010. 

\bibitem[\protect\citeauthoryear{Lim et al.}{2024}]{Limetal2024} Lim G., Im M., Paek G.~S.~H., Yoon S.-C., Imsng Team, 2024, ASPC, 536, 29

\bibitem[\protect\citeauthoryear{Liu, R{\"o}pke, \& Han}{2023}]{Liuetal2023Rev} Liu Z.-W., R{\"o}pke F.~K., Han Z., 2023, RAA, 23, 082001. 


\bibitem[Livio \& Mazzali(2018)]{LivioMazzali2018} Livio, M., \& Mazzali, P.\ 2018, Physics Reports, 736, 1 


\bibitem[Maeda, \& Terada(2016)]{MaedaTerada2016} Maeda, K., \& Terada, Y.\ 2016, International Journal of Modern Physics D, 25, 1630024

\bibitem[\protect\citeauthoryear{Magee et al.}{2025}]{Mageeetal2025} Magee M.~R., Killestein T.~L., Pursiainen M., Godson B., Jarvis D., Jim{\'e}nez-Palau C., Lyman J.~D., et al., 2025, arXiv:2506.02118. 

\bibitem[Maoz et al.(2014)]{Maozetal2014} Maoz, D., Mannucci, F., \& Nelemans, G.\ 2014, \araa, 52, 107


\bibitem[\protect\citeauthoryear{Mehta et al.}{2024}]{Mehtaetal2024} Mehta V., Sullivan J., Fisher R., Ohshiro Y., Yamaguchi H., Bhargava K., Neopane S., 2024, MNRAS, 532, 1087. 

\bibitem[\protect\citeauthoryear{Meng \& Podsiadlowski}{2017}]{MengPodsiadlowski2017} Meng X., Podsiadlowski P., 2017, MNRAS, 469, 4763. 

\bibitem[\protect\citeauthoryear{Michaelis \& Perets}{2025}]{MichaelisPerets2025} Michaelis A., Perets H.~B., 2025, arXiv, arXiv:2507.16907

\bibitem[\protect\citeauthoryear{Mor{\'a}n-Fraile et al.}{2024}]{MoranFraileetal2024} Mor{\'a}n-Fraile J., Holas A., R{\"o}pke F.~K., Pakmor R., Schneider F.~R.~N., 2024, A\&A, 683, A44. 

\bibitem[\protect\citeauthoryear{Neopane et al.}{2022}]{Neopaneetal2022} Neopane S., Bhargava K., Fisher R., Ferrari M., Yoshida S., Toonen S., Bravo E., 2022, ApJ, 925, 92. 

\bibitem[\protect\citeauthoryear{O'Hora et al.}{2025}]{OHoraetal2025} O'Hora J., Ashall C., Shahbandeh M., Hsiao E., Hoeflich P., Stritzinger M.~D., Galbany L., et al., 2025, ApJ, 984, 34. 



\bibitem[\protect\citeauthoryear{Padilla Gonzalez et al.}{2024}]{PadillaGonzalezetal2024} Padilla Gonzalez E., Howell D.~A., Terreran G., McCully C., Newsome M., Burke J., Farah J., et al., 2024, ApJ, 964, 196. 

\bibitem[\protect\citeauthoryear{Palicio et al.}{2024}]{Palicioetal2024} Palicio P.~A., Matteucci F., Della Valle M., Spitoni E., 2024, A\&A, 689, A203. 


\bibitem[\protect\citeauthoryear{Pan, Ruiz-Lapuente, \& Gonz{\'a}lez Hern{\'a}ndez}{2025}]{Panetal2025} Pan K.-C., Ruiz-Lapuente P., Gonz{\'a}lez Hern{\'a}ndez J.~I., 2025, arXiv:2507.01331. 

\bibitem[\protect\citeauthoryear{Pearson et al.}{2024}]{Pearsonetal2024} Pearson J., Sand D.~J., Lundqvist P., Galbany L., Andrews J.~E., Bostroem K.~A., Dong Y., et al., 2024, ApJ, 960, 29. 

\bibitem[\protect\citeauthoryear{Perets et al.}{2019}]{Peretsetal2019} Perets H.~B., Zenati Y., Toonen S., Bobrick A., 2019, arXiv:1910.07532. 

\bibitem[\protect\citeauthoryear{Phillips et al.}{2024}]{Phillipsetal2024} Phillips M.~M., Ashall C., Brown P.~J., Galbany L., Tucker M.~A., Burns C.~R., Contreras C., et al., 2024, ApJS, 273, 16. 

\bibitem[\protect\citeauthoryear{Pollin et al.}{2024}]{Polinetal2024} Pollin J.~M., Sim S.~A., Pakmor R., Callan F.~P., Collins C.~E., Shingles L.~J., R{\"o}pke F.~K., et al., 2024, MNRAS, 533, 3036. 


\bibitem[\protect\citeauthoryear{Pollin et al.}{2025}]{Pollinetal2025} Pollin J.~M., Sim S.~A., Shingles L.~J., Pakmor R., Callan F.~P., Collins C.~E., Roepke F.~K., et al., 2025, arXiv:2507.05000. 

\bibitem[\protect\citeauthoryear{Rajavel, Townsley, \& Shen}{2025}]{Rajaveletal2025} Rajavel N., Townsley D.~M., Shen K.~J., 2025, ApJ, 979, 54. 

\bibitem[\protect\citeauthoryear{Raskin \& Kasen}{2013}]{RaskinKasen2013} Raskin C., Kasen D., 2013, ApJ, 772, 1. 

\bibitem[\protect\citeauthoryear{Ruiter}{2020}]{Ruiter2020} Ruiter A.~J., 2020, IAUS, 357, 1. 

\bibitem[\protect\citeauthoryear{Ruiter \&  Seitenzahl}{2025}]{RuiterSeitenzahl2025} Ruiter A.~J., Seitenzahl I.~R., 2025, A\&ARv, 33, 1. 

\bibitem[Ruiz-Lapuente(2019)]{RuizLapuente2019} Ruiz-Lapuente, P.\ 2019, \nar, 85, 101523

\bibitem[\protect\citeauthoryear{Schinasi-Lemberg \& Kushnir}{2025}]{SchinasiLembergKushnir2025} Schinasi-Lemberg E., Kushnir D., 2025, MNRAS, 536, 3041. 



\bibitem[\protect\citeauthoryear{Sharon \& Kushnir}{2025}]{sharonKushnirWygoda2025}  Sharon A., Kushnir D., Wygoda N., 2025, MNRAS, 540, 3247. 


\bibitem[\protect\citeauthoryear{Sharon, Kushnir, \& Schinasi-Lemberg}{2025}]{Sharonetal2025} Sharon A., Kushnir D., Schinasi-Lemberg E., 2025, MNRAS, 540, 348. 


\bibitem[\protect\citeauthoryear{Shen, Boos, \& Townsley}{2024}]{Shenetal2024} Shen K.~J., Boos S.~J., Townsley D.~M., 2024, ApJ, 975, 127. 

\bibitem[\protect\citeauthoryear{Simotas, Bildsten, \& Prust}{2025}]{Simotasetal2025} Simotas K., Bildsten L., Prust L.~J., 2025, arXiv:2507.06412. 

\bibitem[\protect\citeauthoryear{Soker}{2018}]{Soker2018Rev} Soker N., 2018, SCPMA, 61, 49502. 

\bibitem[\protect\citeauthoryear{Soker}{2019a}]{Soker2019Rev} Soker N., 2019a, NewAR, 87, 101535. 

\bibitem[\protect\citeauthoryear{Soker}{2019b}]{Soker2019b} Soker N., 2019b, MNRAS, 490, 2430. 

\bibitem[\protect\citeauthoryear{Soker}{2022}]{Soker2022RAA} Soker N., 2022, RAA, 22, 035025. 

\bibitem[\protect\citeauthoryear{Soker}{2024a}]{Soker2024Rev} Soker N., 2024a, OJAp, 7, 31. 

\bibitem[\protect\citeauthoryear{Soker}{2024b}]{Soker2024RAAPN} Soker N., 2024b, RAA, 24, 015012. 


\bibitem[\protect\citeauthoryear{Soker}{2024c}]{Soker2024Comment} Soker N., 2024c, arXiv:2406.07363. 

\bibitem[\protect\citeauthoryear{Soker}{2025}]{Soker2025SNR0509} Soker N., 2025, OJAp, 8, 36. 


\bibitem[\protect\citeauthoryear{Soker \& Bear}{2023}]{SokerBear2023} Soker N., Bear E., 2023, MNRAS, 521, 4561. 


\bibitem[\protect\citeauthoryear{Soker et al.}{2013}]{Sokeretal2013} Soker N., Kashi A., Garc{\'\i}a-Berro E., Torres S., Camacho J., 2013, MNRAS, 431, 1541. 

\bibitem[\protect\citeauthoryear{Terwel et al.}{2025a}]{Terweletal2025a} Terwel J.~H., Maguire K., Dimitriadis G., Smith M., Reusch S., Lacroix L., Galbany L., et al., 2025a, A\&A, 694, A11. 

\bibitem[\protect\citeauthoryear{Terwel et al.}{2025b}]{Terweletal2025b} Terwel J.~H., Maguire K., Sollerman J., Wiseman P., Chen T.~X., Graham M.~J., Laher R.~R., et al., 2025b, A\&A, 697, A143. 


\bibitem[\protect\citeauthoryear{Tsalapatas et al.}{2025}]{Tsalapatasetal2025} Tsalapatas K., Sollerman J., Chiba R., Kool E., Johansson J., Rosswog S., Schulze S., et al., 2025, arXiv:2507.08532. 

\bibitem[\protect\citeauthoryear{Uchida et al.}{2024}]{Uchidaetal2024} Uchida H., Kasuga T., Maeda K., Lee S.-H., Tanaka T., Bamba A., 2024, ApJ, 962, 159. 


\bibitem[\protect\citeauthoryear{Vink{\'o}, Szalai, \& K{\"o}nyves-T{\'o}th}{2023}]{Vinkoetal2023} Vink{\'o} J., Szalai T., K{\"o}nyves-T{\'o}th R., 2023, Univ, 9, 244. 

\bibitem[\protect\citeauthoryear{Wang}{2018}]{Wang2018} Wang B., 2018, RAA, 18, 049. 

  
\bibitem[\protect\citeauthoryear{Wang et al.}{2024}]{Wangetal2024} Wang Q., Rest A., Dimitriadis G., Ridden-Harper R., Siebert M.~R., Magee M., Angus C.~R., et al., 2024, ApJ, 962, 17. 


\bibitem[\protect\citeauthoryear{Wang \& Meng}{2025}]{WangMeng2025} Wang X., Meng X., 2025, A\&A, 699, A35. 

\bibitem[\protect\citeauthoryear{Wang, Chen, \& Pan}{2025}]{WangChenPan2025} Wang Y.-H., Chen H.-P., Pan K.-C., 2025, ApJ, 989, 72. 


\bibitem[\protect\citeauthoryear{Wu et al.}{2025}]{Wuetal2025} Wu W., Jiang J.-. an ., Meng D., Xu Z., Maeda K., Doi M., Nomoto K., et al., 2025, arXiv, arXiv:2507.15609

 
\bibitem[\protect\citeauthoryear{Yang, Thomas Tam, \& Yang}{2022}]{Yangetal2022} Yang H.-W., Thomas Tam P.-H., Yang L., 2022, RAA, 22, 105014. 

\bibitem[\protect\citeauthoryear{Zenati et al.}{2023}]{Zenatietal2023} Zenati Y., Perets H.~B., Dessart L., Jacobson-Gal{\'a}n W.~V., Toonen S., Rest A., 2023, ApJ, 944, 22. 

\bibitem[\protect\citeauthoryear{Zenati, Toonen, \& Perets}{2019}]{Zenatietal2019} Zenati Y., Toonen S., Perets H.~B., 2019, MNRAS, 482, 1135. 

\bibitem[\protect\citeauthoryear{Zingale et al.}{2024}]{Zingaleetal2024} Zingale M., Chen Z., Rasmussen M., Polin A., Katz M., Smith Clark A., Johnson E.~T., 2024, ApJ, 966, 150. 


\end{thebibliography}
\end{document}